\documentclass[aps,twocolumn,floats,nofootinbib,groupedaddress]{revtex4-1}

\usepackage{graphicx}
\usepackage{amsmath}

\begin{document}

\title{Predictability and hierarchy in \emph{Drosophila} behavior}

\author{Gordon J. Berman}
\email{E-mail: gordon.berman@emory.edu.  Current address: Department of Biology, Emory University, Atlanta, GA 30322}
\author{William Bialek} 
\author{Joshua W. Shaevitz}
\affiliation{Joseph Henry Laboratories of Physics and Lewis--Sigler Institute for Integrative Genomics, Princeton University, Princeton, NJ 08544}

\date{\today}

\begin{abstract}
Even the simplest of animals exhibit behavioral sequences with complex temporal dynamics.  Prominent amongst the proposed organizing principles for these dynamics has been the idea of a hierarchy, wherein the movements an animal makes can be understood as a set of nested sub-clusters. Although this type of organization holds potential advantages in terms of motion control and neural circuitry, measurements demonstrating this for an animal's entire behavioral repertoire have been limited in scope and temporal complexity. Here, we use a recently developed unsupervised technique to discover and track the occurrence of all  stereotyped behaviors performed by fruit flies moving in a shallow arena.  Calculating the optimally predictive representation of the fly's future behaviors, we show that fly behavior exhibits multiple time scales and is organized into a hierarchical structure that is indicative of its underlying behavioral programs and its changing internal states.
\end{abstract}

\maketitle

\section{Introduction}

Animals perform a vast array of behaviors as they go about their daily lives, often in what appear to be repeated and non-random patterns.  These sequences of actions, some innate and some learned, have dramatic consequences with respect to survival and reproductive function--from feeding, grooming, and locomotion to mating, child rearing, and the establishment of social structures.  Moreover, these patterns of movement can be viewed as the final output of the complicated interactions between an organism's genes, metabolism, and neural signaling.  As a result, understanding the principles behind how an animal generates behavioral sequences can provide a window into the biological mechanisms underlying the animal's movements, appetites, and interactions with its environment, as well as broader insights into how behaviors evolve.

The prevailing theory for the temporal organization of behavior, rooted in work from neuroscience, psychology, and evolution, is that the pattern of actions performed by animals is hierarchical \cite{tinbergen51,dawkins76,simon73}. In such a framework, actions are nested into modules on many scales, from simple motion primitives to complex behaviors to sequences of actions.  
Neural architectures related to behavior, such as the motor cortex, are anatomically hierarchical,  supporting the idea that animals use a hierarchical representation of behavior in the brain \cite{Graziano:2007gm,Bassett:2008ij,Dombeck:2009cd,Chen:2012dy}.  Additionally, hierarchical organization is a hallmark of human design, from the layout of cities to the wiring of the internet, and its potential use in various biological contexts has been proposed as an organizing principle \cite{dawkins76}. 

Despite the theoretical attractiveness of behavioral hierarchy, measurements showing that a particular animal's behavioral repertoire is organized in this manner often are limited in their applicability and scope.  Typically, observations of hierarchy in the ordering of movement have considered a single behavioral type, such as grooming, ignoring relationships between more varied behavioral motifs \cite{DAVIS:1974vm,Dawkins:1976wx,Lefebvre:1981wb,Lefebvre:1982fh,Lefebvre:1982wk,Seeds:2014eu}.  Perhaps more problematic is that most analyses of behavior make use of methods, such as hierarchical clustering, that implicitly or explicitly {\em impose} a hierarchical structure onto the data without showing that such a representation is accurate.  Lastly, to our knowledge, all measurements of a hierarchical organization of behavior limit their analysis to behavioral dynamics at a single time scale.  This scale is often given by the results of fitting a Markov model, where the next step in a behavioral pattern only depends on the animal's current state.  Even in the simplest of animals, however, there are many internal states such as hunger, reproductive drive, etc., and sequences of behaviors possess an effective memory of an animal's behavioral state that persists well into the future, a result noted in a wide variety of systems \cite{Miller:2003ii,Heiligenberg:1973gy,Jin:2011fw,Dawkins:1973ta}. 

In this paper, we study the behavioral repertoire of fruit flies (\emph{Drosophila melanogaster}), attempting to characterize the temporal organization of their movements over the course of an hour.  Decomposing the flies' movements into a set of stereotyped behaviors without making any \emph{a priori} behavioral definitions \cite{Berman2014}, we find that their behavior exhibits long time scale,s far beyond what would be predicted from a Markovian model.  Applying methods from information theory, we show that a hierarchical representation of actions optimally predicts the future behavioral state of the fly.  These results show that the best way to understand how  future actions follow from the current behavioral state is to group these current behaviors in a nested manner, with fine grained partitions being useful in predicting the near future, and coarser partitions being sufficient for predicting the relatively distant future.  These results show that these animals control their movement via a hierarchy of behaviors at varying time scales, affirming and making precise a key concept in ethology.

\section{Experiments and behavioral states}
As a testbed for probing questions of behavioral organization and hierarchy, we sought to measure the entire behavioral repertoire of a population of \emph{Drosophila melanogaster} in a specific environmental context. We probed the behavioral repertoire of individual, ground-based fruit flies in a largely featureless circular arena for one hour using a 100Hz camera.  Under these conditions, flies display many complex behaviors, including locomotion and grooming, that involve multiple parts of their bodies interacting at varying time scales. We recorded videos of 59 male flies using a custom-built tracking setup, producing more than 21 million images \cite{Berman2014}.

\begin{figure}
\centering
\includegraphics[width=\columnwidth]{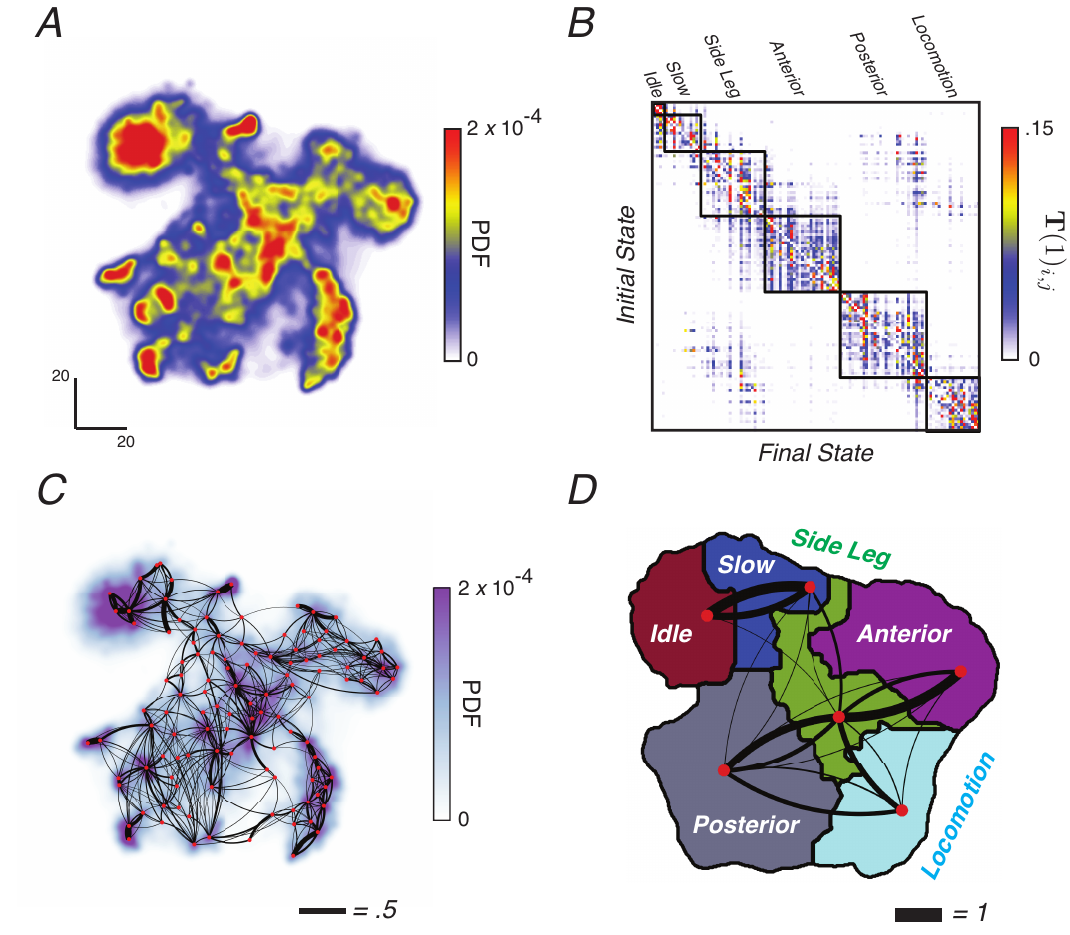}
\caption{Transition probabilities and behavioral modularity.  (A) Behavioral space probability density function (PDF).  Here, each peak in the distribution corresponds to a distinct stereotyped movement.  (B) One-step Markov transition probability matrix $\mathbf{T}(\tau=1)$. The 117 behavioral states are grouped by applying the predictive information bottleneck calculation and allowing 6 clusters (Eq. \ref{IB}). Black lines denote the cluster boundaries.  (C) Transitions rates plotted on the behavioral map.  Each red point represents the maximum of the local PDF, and the black lines represent the transition probabilities between the regions.  Line thicknesses are proportional to the corresponding value of $\mathbf{T}(\tau=1)_{ij}$, and right--handed curvature marks the direction of the transition.  For clarity, all lines representing transition probabilities of less than .05 are omitted.  (D)  The clusters found using the information bottleneck approach (colored regions) are contiguous in the behavioral space. Behavioral labels associated with each partitioned graph cluster from B are shown. Black lines thickness represents the conditional transition probabilities between clusters.  All transition probabilities less than .05 are omitted.}\label{transition_fig} 
\end{figure}

These data were used to generate a two--dimensional map of fly behavior based on an unsupervised approach that automatically identifies stereotyped actions (Fig. \ref{transition_fig}A, for full details see \cite{Berman2014}).  Briefly, this approach takes a set of translationally and rotationally aligned images of the flies and decomposes the dynamics of the observed pixel values into a low--dimensional basis set describing the flies' posture.  Time series are produced by projecting the original pixel values onto this basis set and the local spectrogram of these trajectories is then embedded into two dimensions \cite{vanderMaaten:2008tm}. Each position in the behavioral map corresponds to a unique set of postural dynamics; although this was not required by the analysis, nearby points represent similar motions, i.e. those involving related body parts executing similar temporal patterns.

In the resulting behavioral space, $\mathbf{z}$, we estimate the probability distribution function $P(\mathbf{z})$ and find that it contains a set of peaks corresponding to  short segments of movement that are revisited multiple times by multiple individuals (Figure \ref{transition_fig}A). Pauses in the trajectories through this space, $\mathbf{z}(t)$, are interspersed with quick movements between the peaks.  These pauses in $\mathbf{z}(t)$ at a particular peak correspond to the fly performing one of a large set of distinct, stereotyped behaviors such as right wing grooming, proboscis extension, or alternating tripod locomotion \cite{Berman2014}.  In all, we identify 117 unique stereotyped actions, with similar behaviors, i.e. those that utilize similar body parts at similar frequencies, located near each other in the behavioral map. A watershed algorithm is used to separate the peaks and, combined with a threshold on $d\mathbf{z}(t)/dt$, to segment each movie into a sequence of discreet, stereotyped behaviors.

In this paper, we treat pauses at these peaks to be our states, the lowest level of description of  behavioral organization, and investigate the pattern of behavioral transitions among these states over time.  We count time in units of the transitions between states, so we have a description of behavior as a discrete variable $S(n)$ that can take on $N = 117$ different values at each discrete time $n$. Note that since we count time in units of transitions, we always have $S(n+1) \ne S(n)$.  Combining data from all 59 flies, we observe $\approx 6.4\times 10^5$ behavioral transitions, or  $\approx 10^4$ per experiment.

\begin{figure*}
\centering
\includegraphics[width=1.5\columnwidth]{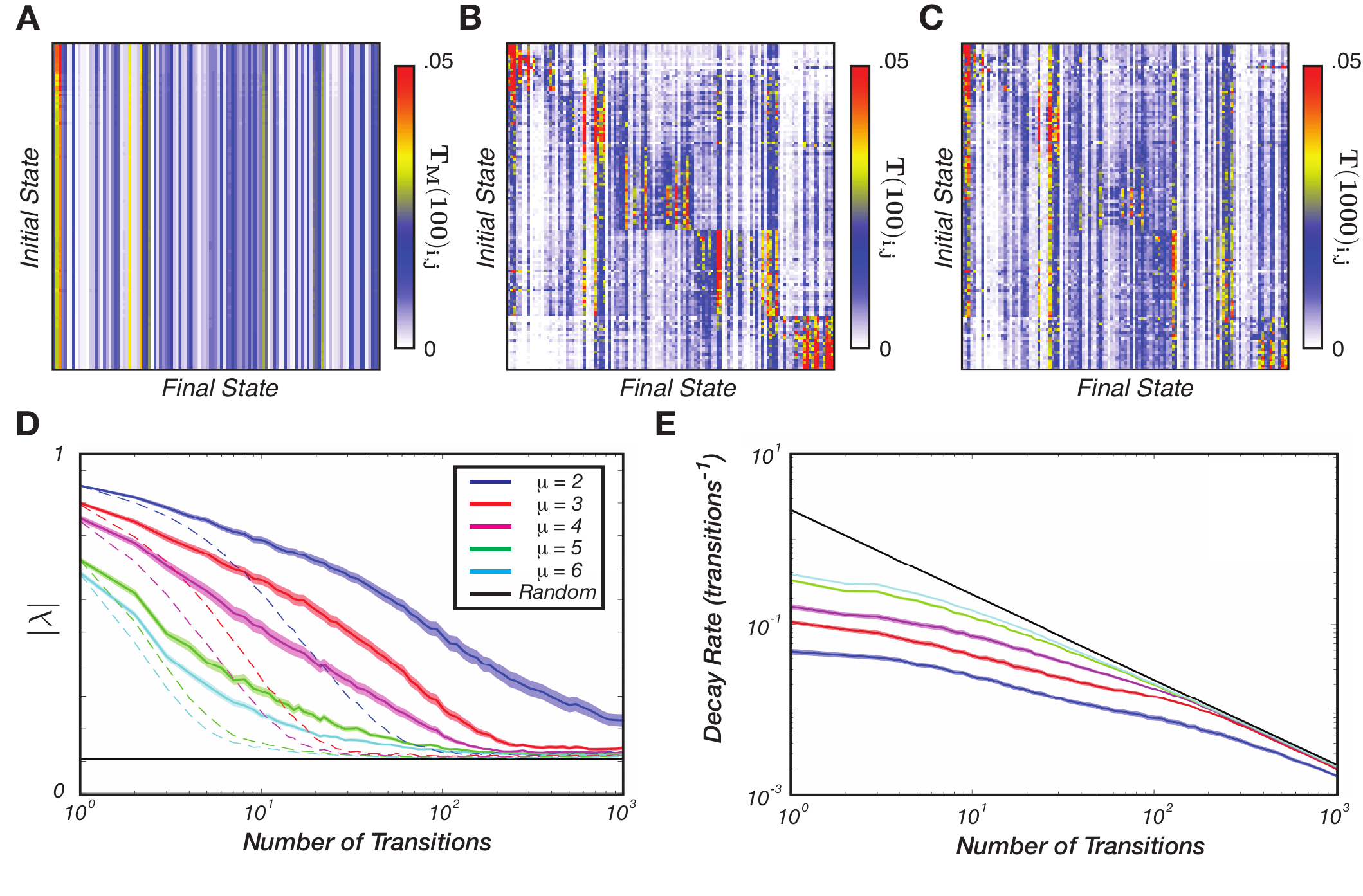}
\caption{Long time scale transition matrices and non--Markovian dynamics.  (A) Markov model transition matrix for $\tau=100$, $\mathbf{T}_M(100)$, from Eq (\ref{TM}).  (B and C) Transition matrices for $\tau=100$ and $\tau=1,000$, respectively, from Eq (\ref{Tij}).  (D) Absolute values of the leading eigenvalues  of the transition matrices $\mathbf{T}{(\tau)}$ as a function of $\tau$.  The curves represent the average over all flies, and thicknesses represent the standard error of the mean.  Dashed lines are the predictions for the Markov model $\mathbf{T}_M(\tau )$.  The black line is a noise floor, corresponding to the typical value of the second largest eigenvalue in a transition matrix calculated from random temporal shuffling of our finite data set.  (E) Eigenmode decay rates, $r_\mu(\tau) \equiv -\log \vert\lambda_\mu (\tau)\vert / \tau$, as a function of the number of transitions.  Line colors represent the same modes as in (D) and the black line again corresponds to a ``noise floor,'' in this case the largest decay rate that we can resolve above the random structures present in our finite sample.}\label{time_scale_fig}
\end{figure*}

\section{Transition matrices and non-Markovian time scales}
To investigate the temporal pattern of behaviors, we first calculated the behavioral transition matrix over different time scales, 
\begin{equation}
	\left[\mathbf{T}(\tau)\right]_{i,j} \equiv p(S(n+\tau) = i \vert S(n) = j),  \label{Tij}
\end{equation}
which describes the probability that the animal will go from state $j$ to state $i$ after $\tau$ transition steps. We expect that this distribution becomes less and less structured as $\tau$ increases because we lose the ability to make predictions of the future state as the horizon of our predictions extends further.  In addition, it will be useful to think about these matrices in terms of their eigendecompositions: 
\begin{equation}
\left[\mathbf{T}(\tau)\right]_{i,j} = \sum_\mu \lambda_\mu (\tau) u_i^\mu (\tau) v_j^\mu (\tau) ,
\end{equation}
where $ \mathbf{u}^\mu \equiv \{u_i^\mu\} $ and $\mathbf{v}^\mu  \equiv \{v_i^\mu\} $ are the left and right eigenvectors, respectively, and $\lambda_\mu (\tau)$ is the eigenvalue with the $\mu^{\rm{th}}$ largest modulus.  Because probability is conserved in the transitions, the largest eigenvalue will always be equal to one, $\lambda_1(\tau)=1$, and $v_i^1(\tau)$ describes the stationary distribution over states at long times.  All the other eigenvalues have magnitudes less than one, $\vert\lambda_{\mu\neq 1}(\tau)\vert < 1$, and describe the loss of predictability over time, as shown in more detail below.

The matrix $\mathbf{T}(\tau = 1)$ describes the probability of transitions from one state to the next, the most elementary steps of behavior  (Fig. \ref{transition_fig}B).  To the eye, this transition matrix appears modular, with most transitions out of any given state only going to one of a handful of other states. By appropriately organizing the states in Figure \ref{transition_fig}B, $\mathbf{T}(\tau=1)$ takes on a nearly block-diagonal structure, which can be broken up into modular clusters using the information bottleneck formalism (see below).  Plotting this matrix on the behavioral map itself (Fig. \ref{transition_fig}C), we see that the transitions are largely localized, with nearly all large probability transitions occurring between nearby behaviors. Furthermore, the transition clusters are contiguous in the behavioral space, defining gross categories of motion including locomotion, behaviors involving anterior parts of the body etc. (Fig. \ref{transition_fig}D). 

It is important to note that $\mathbf{T}{(\tau=1)}$ does not directly contain information about the  location of behavioral states in the two dimensional map, and hence any relationship we observe between the transition structure and the patterning of behaviors in the map is a consequence of the animal's behavior and not the way we construct the analysis. We thus conclude that behavioral transitions are mostly restricted to occur between similar actions---e.g.,  grooming behaviors are typically followed by other grooming behaviors of close-by body parts and animals transition between locomotion gates systematically by changing gate speed and velocity.  These observations are consistent with classical ideas of postural facilitation and previous observations that transitions largely occur between similar behaviors \cite{Dawkins:1976wx,Takahata:1981wd,Ackermann:1991jx,Hopkins:2002ji}.

We begin to see the necessity of looking at longer time scales as we measure the transition matrices for $\tau \gg 1$.  If the observed dynamics are purely Markovian, then the transitions from one state to the next do not depend on the history of behavior, and $\mathbf{T}(\tau =1)$ provides a complete characterization of the system.  In particular, if the behavior is Markovian then we can calculate the transition matrix after $\tau$ state just by iterating the matrix from one step:
\begin{equation}
\mathbf{T}_M{(\tau)} \equiv \left[\mathbf{T}(1)\right] ^\tau = \sum_\mu [\lambda_\mu(1)]^\tau \mathbf{u}_\mu (1) \mathbf{v}_\mu(1). \label{TM}
\end{equation}
Because $\vert\lambda_\mu(1)\vert < 1$ for all but the leading eigenvalue, the contributions from the $\mu>1$ terms decay to zero exponentially as $\tau \to \infty$. For very long times, therefore,  $\mathbf{T}_M{(\tau)}$ loses all information about the current state and instead reflects the average probabilities of performing any particular behavior. Thus, in a Markovian system, the slowest time scale in the system is determined by $\vert\lambda_2(1)\vert$, resulting in a characteristic decay time $t_2 = -1 / \log \vert\lambda_2(1)\vert$.  Calculating these eigenvalues for each fly and averaging, we find $\langle \lambda_2(1)\rangle = 0.953\pm 0.004$, or $\langle t_2 \rangle = 29\pm 2$ transitions.  Thus, any memory that extends beyond $\approx 30$ transitions into the future is direct evidence for hidden states that carry a memory over longer times and modulate behavior.

Initial evidence for long-time structure in $\mathbf{T}(\tau)$ comes by comparing the lack of structure within $\mathbf{T}_M(100)$ to that within $\mathbf{T}{(\tau)}$ for $\tau = 100$ and $\tau=1,000$ (Fig \ref{time_scale_fig}A-C).  After 100 transitions, ($\approx 3\langle t_2 \rangle $), the Markov model retains essentially no information, as demonstrated by the similarity between all of the rows, implying that all transitions have been randomized.  Conversely, although some of the block--diagonal structure from Fig. \ref{transition_fig}B has dissipated, we see that $\mathbf{T}{(100)}$ and $\mathbf{T}{(1000)}$ retrain a great deal of non-randomness.

This observation can be made more precise by looking at the eigenvalue spectra of the transition matrices.  In Figure \ref{time_scale_fig}D, we plot $\vert \lambda_\mu(\tau) \vert$ as a function of $\tau$ for $\mu=2$ through 6 (solid color lines) in addition to the predictions from the Markov model of Eq (\ref{TM}) based on $\mathbf{T}(1)$ (colored dashed lines).  In a Markovian system, it would be more natural to plot these results with a logarithmic axis for $\lambda$, but here we see that structure extends over such a wide range of time scales that we need a logarithmic axis for $\tau$.  We can make this difference more obvious by measuring the apparent decay rate, $r_\mu(\tau) = -\log \vert\lambda_\mu (\tau)\vert / \tau$, which should be constant for a Markovian system.  For the leading mode, the apparent decay rate falls by nearly two orders of magnitude before the corresponding eigenvalue is lost in the noise (Figure \ref{time_scale_fig}E).  Similar patterns appear in higher modes, but we have more limited dynamic range for observing them.

These results are direct evidence that many time scales are required to model behavioral sequences, even in this simple context where no external stimuli are provided.  Accordingly, we can infer that the organism must have internal states that we do not directly observe, even though we are making rather thorough measurements of the motor output. Roughly speaking, the appearance of decay rates $\approx 10^{-3}$ means that the internal states must hold memory across at least $\approx 10^3$ behavioral transitions, or approximately 20 minutes---much longer than any time scale apparent in the Markov model.

\section{Predictability and hierarchy}
The modular structure of the flies' transition matrix, combined with the observed long time scales of behavioral sequences, suggests that we might be able to group the behavioral states into clusters that preserve much of the  information that the current behavioral state provides about future actions (predictive information \cite{bialek+al_01}).  Furthermore, we should be able to probe whether this results in a hierarchical organization:  if the states are grouped into a hierarchy, then increasing the number of clusters will largely subdivide existing clusters rather than mix behaviors from two different clusters.

\begin{figure}[b]
\centering
\includegraphics[width=\columnwidth]{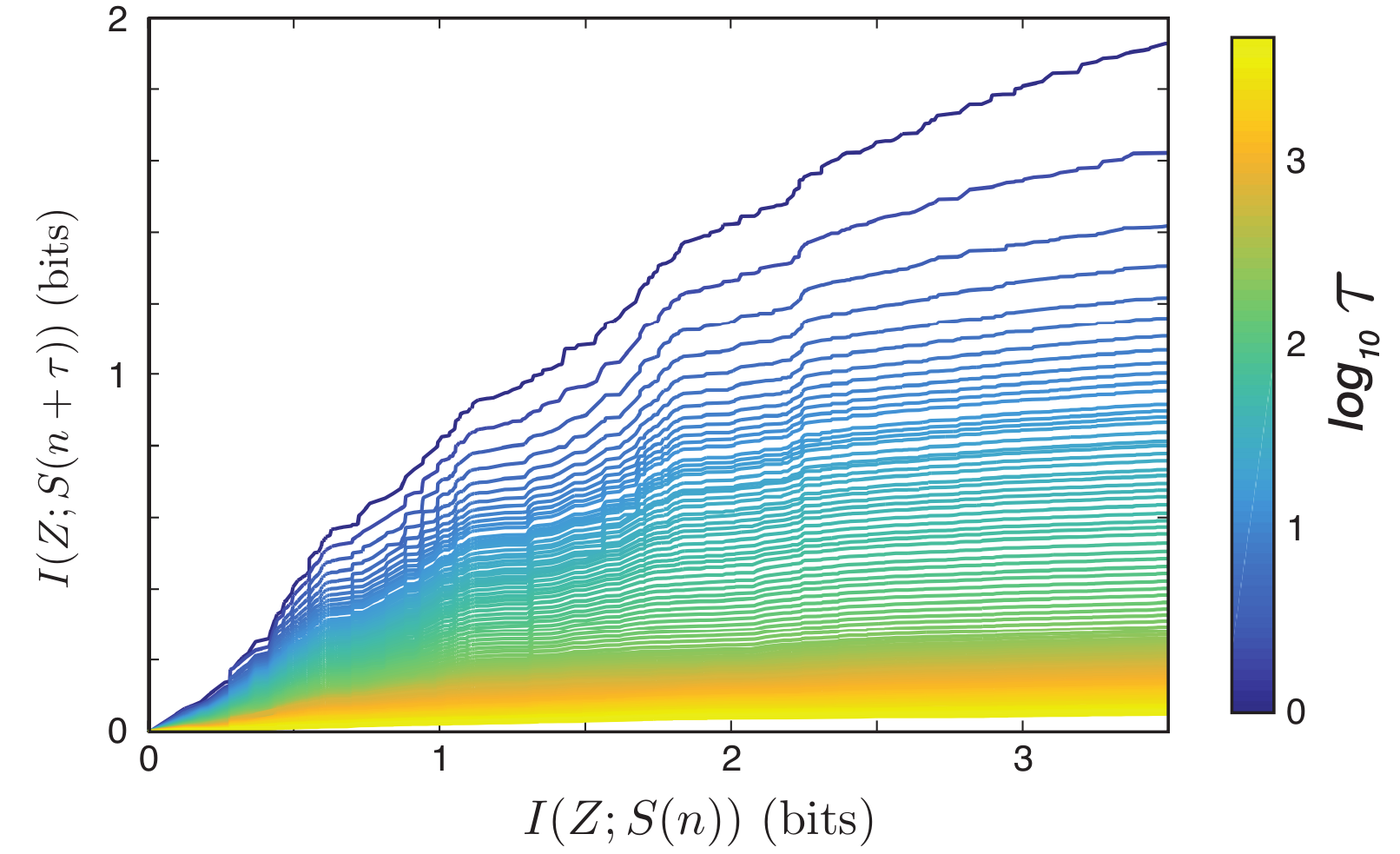}
\caption{Optimal trade-off curves for lags from $\tau=1$ to $\tau=5000$. For each time lag $\tau$, number of  clusters, and $\beta$, we optimize Equation \ref{IB} and plot the resulting complexity of the partitioning, $I(Z; S(n))$, versus the predictive information, $I(Z; S(n+\tau))$.}
\label{trade_off_curves}
\end{figure}

\begin{figure*}
\centering
\includegraphics[width=1.5\columnwidth]{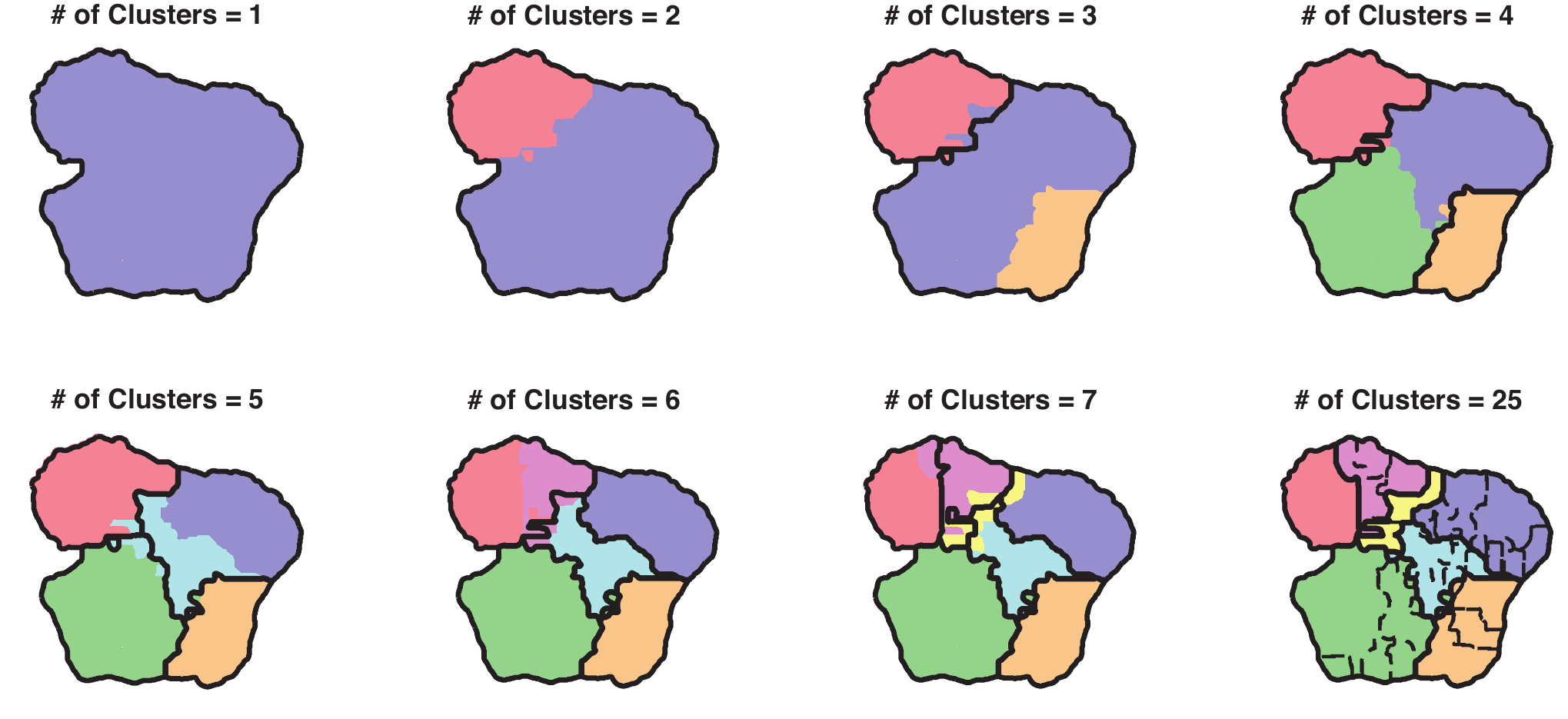}
\caption{Information bottleneck partitioning of behavioral space for $\tau = 67$ (approximately twice the longest time scale in the Markov model).  Borders from the previous partitions are shown in black.  For 25 clusters (bottom right), the partitions, still contiguous, are denoted by dashed lines.}\label{bottleneck_hierarchy}
\end{figure*}

To make this idea more precise, we hope to map the behaviors into groups, $S(n) \rightarrow Z$, that compress our description in a way that preserves information about a state $\tau$ transitions in the future, $S(n+\tau)$. Mathematically, this means that we should maximize the information about the future, $I(Z; S(n+\tau))$, while holding fixed the information that we keep about the past, $I(Z; S(n))$.   Introducing a Lagrange multiplier to hold $I(Z; S(n))$ fixed, we wish to maximize
\begin{equation}
{\cal F} = I(Z; S(n+\tau)) - \beta I(Z; S(n)).
\label{IB}
\end{equation}
At $\beta  = 0$ we retain the full complexity of the 117 behavioral states, and as we increase $\beta$, we are forced to tighten our description into a more and more compressed form, thus losing predictive power.  This is an example of the information bottleneck problem  \cite{tishby99}.  If the compressed description $Z$ involves a fixed number of clusters, then we find solutions that range from soft clustering, where behaviors can be assigned to more than one cluster probabilistically, to hard clustering, where each behavior belongs to only one cluster,  as $\beta$ increases; changing the number of clusters allows us to move along a curve that trades complexity of description against predictive power, as shown in Fig \ref{trade_off_curves}  (see \S\ref{IB_methods} for details).

As expected, the optimal curves move downward as the time lag increases, implying that the ability to predict the behavioral state of the animal decreases as we look further into the future.  We also observe a relatively rapid decrease in the height of these curves for small $\tau$, followed by increasingly-closely spaced optimal curves as the lag length increases. It this slowing that is indicative of the long time-scales in behavior.

\begin{figure}[b]
\centering
\includegraphics[width=\columnwidth]{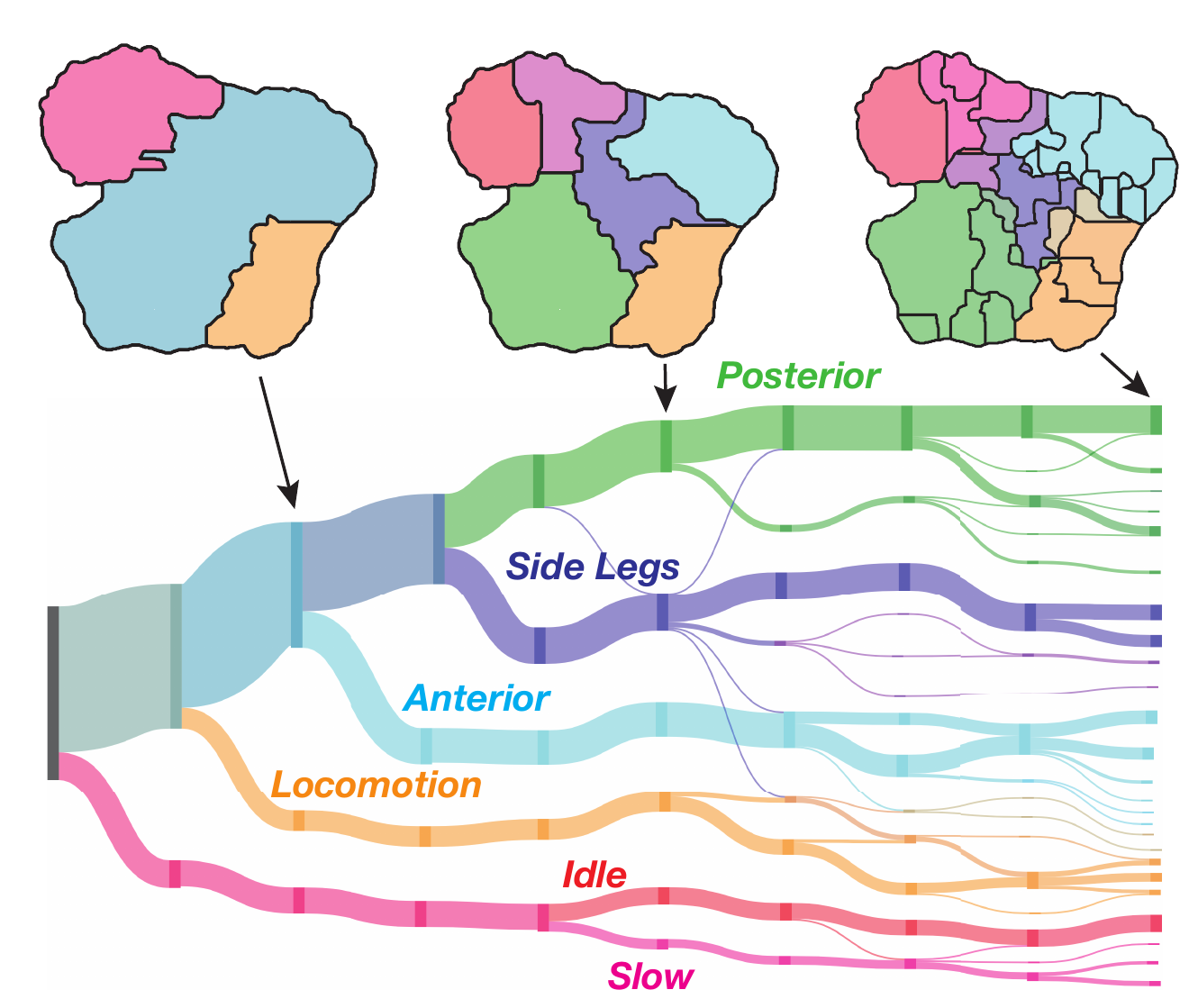}
\caption{Hierarchical organization for optimal solutions with lag $\tau=100$ ranging from 1 cluster to 25.  The displayed clusterings are those that have the largest value of $I(Z;S(n+\tau))$ for that number of clusters.  The length of the vertical bars are proportional to the percentage of time a fly spends in each of the clusters, and the lines flowing horizontally from left to right are proportional in thickness to the flux from the clustering on the left to the clustering on the right.  Fluxes less than .01 are suppressed for clarity.}\label{t_100_flow}
\end{figure}

Along each of these trade-off curves lie partitions of the behavioral space that contain an increasing number of clusters. We can make several observations about these data. First, in agreement with our investigation of the single-step transition matrix, we find that the clusters are spatially contiguous in the behavioral map as exemplified in Figure \ref{bottleneck_hierarchy} for $\tau=67$. Thus, even when we add in the long time-scale dynamics, we find that transitions  predominantly occur between similar behaviors. Second, these spatially-contiguous clusters separate hierarchically as we increase the number of clusters, i.e. new clusters largely result from subdividing existing clusters instead of emerging from multiple existing clusters.  One example of this can be seen in Figure \ref{t_100_flow}, where the probability flow between partitions of increasing size subdivide in a tree-like manner.  It is important to note that these results are not built in to the information bottleneck algorithm:  we can solve the bottleneck problem for different numbers of clusters independently, and hence (in contrast to hierarchical clustering) this method could have found non-hierarchical evolution with new clusters comprised of behaviors from many other clusters, That this does not happen is strong evidence that fly behavior is  organized hierarchically.

\begin{figure}
\centering
\includegraphics[width=\columnwidth]{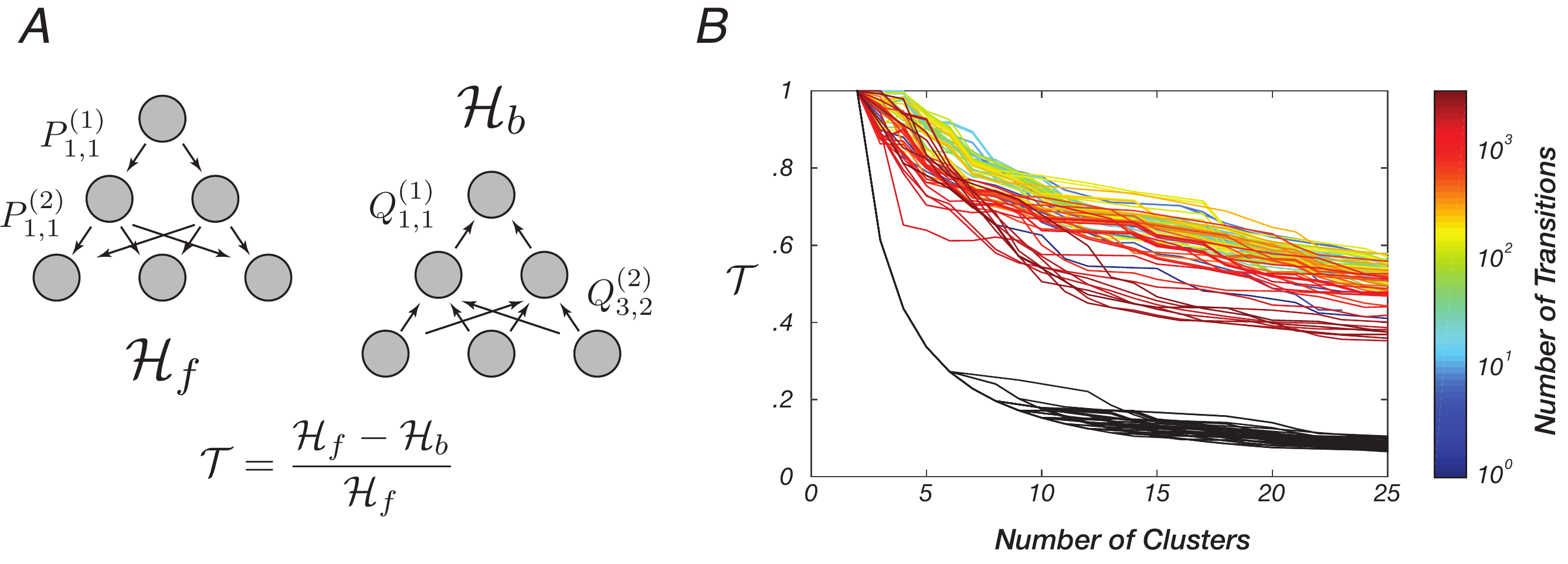}
\caption{Partitionings are tree-like over all measured time scales.  (A) Definition of the treeness metric, $\mathcal{T}$;  Methods for details.  (B) $\mathcal{T}$ as a function of the number of transitions in the future and the number of clusters in the most fine-grained partition.  Colored lines represent values of $\mathcal{T}$ for partitions at varying times in the future, and black lines are values for randomized graphs generated from partitionings that were assigned randomly. \label{treeness_index_fig}}
\end{figure}

We can go beyond this qualitative description, however, by quantifying the degree of hierarchy in our representation as the number of clusters increases using a ``treeness'' metric, $\mathcal{T}$ (Fig. \ref{treeness_index_fig}).  The idea behind this metric, which is similar to the one introduced by Corominas--Murta et al~\cite{CorominasMurtra:2011bw}, is that if our representation is perfectly hierarchical, then each cluster  has precisely one ``parent" in a partitioning with a smaller number of clusters.  Thus, the better our ability to distinguish the lineage of a cluster as it splits through increasingly complex partitionings implies a higher value of $\mathcal{T}$.  More precisely, the treeness index is given by the relative reduction in entropy going backwards rather than forwards through the tree, 
\begin{equation}\label{tree_eqn}
\mathcal{T} = \frac{\mathcal{H}_f-\mathcal{H}_b}{\mathcal{H}_f},
\end{equation}
where $\mathcal{H}_f$ and $\mathcal{H}_b$  are the entropies over all possible paths going forward and backwards, respectively. This metric is bounded between zero and one, $0\le\mathcal{T}\le 1$, and $\mathcal{T} = 1$ implies a perfect hierarchy.

We find that the partitionings derived from the information bottleneck algorithm are much more tree-like than random partitions of the behavioral space (Fig. \ref{treeness_index_fig}B).  This is true even when we attempt to optimally predict behavioral states thousands of transitions into the future.  Thus, by finding optimally-predictive representations that best explain the relationship between states over long time scales, we have uncovered a hierarchical ordering of actions, supporting decades-old theory without relying on hierarchical clustering, Markov models, or limiting the measured behavioral repertoire.

\section{Conclusions}
We have measured the behavioral repertoires for dozens of fruit flies, paying particular attention to the structure of their behavioral transitions.  We find that these transitions exhibit multiple time scales and possess memory that persists thousands of transitions into the future, indicative of internal states that carry memory across thousands of observable behavioral transitions.  Using an information bottleneck approach to find the compressed representations that optimally predict our observed dynamics, we find that behaviors are organized in a hierarchical fashion, with fine grained representations being able to predict short--time structure and coarser representations being sufficient to predict the fly's actions that are further removed in time.  This is fundamentally different from previous measurements of hierarchy in behavior, which were more limited in the types of behaviors they measured, the time scales over which the hierarchy was modeled, and/or relied on hierarchical clustering and other types of analyses that only yield hierarchical outputs.

The type of organization we observe is reminiscent of the functional clustering seen in mouse and primate motor cortex, where groupings of neurons from millimeter scales down to single cells have been found to exhibit increasing temporal correlation as the distance between them decreases \cite{Graziano:2007gm,Dombeck:2009cd}.  Although no such pattern has been specifically found in \emph{Drosophila}, our results suggest that such neuronal patterns may exist. As circuits for different behavioral modules are uncovered, our results suggest that such hierarchical neuroanatomical organization will also be found in the fly, serving as a general principle that may apply across organisms to provide insight towards how the brain controls behavior and adapts to a complex environment.

\begin{acknowledgments}
We thank Ugne Klibaite, David Schwab, and Thibaud Taillefumier for discussions and suggestions. JWS and GJB also acknowledge the Aspen Center for Physics, where many ideas for this work were formulated. This work was funded through awards from the National Institutes of Health (GM098090, GM071508), The National Science Foundation (PHY-1305525, PHY-1451171, CCF-0939370), the Swartz Foundation, and the Simons Foundation.
\end{acknowledgments}

\section{Methods}

\subsection{Experiments} 
We imaged 59 individual male flies (\emph{D. melanogaster}, Oregon-R strain) for an hour each, following the protocols originally described in \cite{Berman2014}.  All flies were within the first two weeks post-eclosion during the filming session.  Flies were placed into the arena via aspiration and were subsequently allowed 5 minutes for adaptation before data collection. All recording occurred between the hours of 9:00 AM and 1:00 PM.  The temperature during all recordings was $25^o \pm 1^o C$.  

\subsection{Generating Markovian Models}
Markovian model data sets were generated by first randomly selecting a state, and then finding another, randomly chosen, instance in the measured data set where the fly was performing that behavior.  The behavior performed immediately after that behavior is chosen, and the process is iterated until the generated sequence is equivalent in size to the original data set, similar to the first-order alphabets generated in Shannon's original work on information theory \cite{Shannon:1948kl}.

\subsection{Predictive Information Bottleneck}\label{IB_methods}
The solution to the information bottleneck problem, Eq (\ref{IB}), obeys a set of self--consistent equations that can be iterated in a manner equivalent to the  Blahut-Arimoto algorithm in rate--distortion theory \cite{blahut72,tishby99}.  For a given $\vert Z \vert = K$ and inverse temperature $\beta$, a random initial condition for $p(z\vert x)$ is chosen, and the following self--consistent equations are iterated until the convergence criterion ($(\mathcal{F}_{t}-\mathcal{F}_{t+1})/\mathcal{F}_t < 10^{-6}$) is met:
\begin{eqnarray}
p(z\vert x) &=& \frac{p(z)}{\mathcal{Z}(\beta,x)}\exp \Bigr[-\beta D_{KL}\Big( p(y\vert x) \vert\vert p(y\vert z) \Big) \Bigr], \\
p(z) &=& \sum_x p(z\vert x) p(x) \\
p(y\vert z) &=& p(y\vert x)p(z\vert x)p(x),
\end{eqnarray}
where $x\in S(n)$, $y\in S(n+\tau)$, $z\in Z$, $D_{KL}$ is the Kullback-Leibler divergence between two probability distributions, and $\mathcal{Z}(\beta,x)$ is a normalizing function.  

Because this study focuses on hard clusterings of the behavioral space, we find solutions by starting at $\beta=0.1$ and annealing with 40 exponentially-spaced values up to $\beta = 500$.  After starting from a random initial condition at the initial value of $\beta$, the optimization is performed at that value until the convergence criterion is met, and that solution is used as the initial condition for the next value of $\beta$.  All intermediate solutions, $p^{(n)}_{\ell}(z\vert x)$ are stored so they can potentially be included in the found Pareto front.  In addition, we perform 24 replicates of this process with different random initial conditions for $K=2,\ldots , 25$ and for 81 time lag values between $n=1$ and $n=5,000$.  

Given the set of solutions for a given lag, we first take the deterministic limit of each clustering ($p(z\vert x) = \delta_{z,\arg\max_{z'} p(z'\vert x)}$) and recalculate $I(Z;S(n))$ and $I(Z;S(n+\tau))$ accordingly.  We then defined the Pareto front, $\xi^{(n)}$, as the set of all solutions, $p^{(n)}_{\ell}(z\vert x)$, such that no other solution for that given lag results in a smaller value for $I(Z;S(n))$ and a larger value for $I(Z;S(n+\tau))$.  Between 150 and 350 solutions were found for all of the fronts.  We choosing a clustering for a fixed number of clusters, here, we always pick the representation along the optimal front that has the highest value of $I(Z;S(n+\tau))$.

\subsection{Treeness Index}
To calculate the treeness index, $\mathcal{T}$, we construct a directed, acyclic forward graph that connects the partitions as the number of clusters increases for a given time lag with values $P^{(\ell )}_{ij}$.  These values are the probability that a state contained in one cluster, $i$, in the partitioning with $\ell$ clusters also belongs to cluster $j$ in the partitioning with $\ell+1$ clusters.  Similarly, we can create the backwards graph, $Q^{(\ell )}_{ij}$, that links clusters in the opposite direction; $Q^{(\ell )}_{ij}$ is the probability that a state in cluster $i$ in the partitioning with $\ell+1$ clusters also belongs to the cluster $j$ in the partitioning containing $\ell$ clusters.

Given these two graphs, we can calculate the entropy of picking a path, $\pi^{(f)}$ in the forward direction versus the entropy of picking a path, $\pi^{(b)}$ in the backwards direction.  These probabilities can be calculated via $p(\pi_{\mathbf{v}}^{(f)})=\prod_{\ell=1}^{N-1} P^{(\ell )}_{v_\ell,v_{\ell+1}}$ and p($\pi^{(b)}_{\mathbf{v}})=\prod_{\ell=1}^{N-1} Q^{(\ell )}_{v_{\ell+1},v_\ell}$, with $\mathbf{v}$ being a chosen sequence of clusters.  Thus, we define the forward and backwards entropies as follows:
\begin{eqnarray}
\mathcal{H}_f &=& -\sum_\mathbf{v\in V} p(\pi_{\mathbf{v}}^{(f)}) \log p(\pi_{\mathbf{v}}^{(f)}) \\ 
\mathcal{H}_b &=& \langle-\sum_\mathbf{w\in W_r} p(\pi_{\mathbf{w}}^{(b)}) \log p(\pi_{\mathbf{w}}^{(b)})\rangle_r  ,
\end{eqnarray}
where $\mathbf{V}$ is the set of all possible paths and $\mathbf{W_r}$ is the set of all paths ending at cluster $r$ in the most fine-grained partitioning.  $\langle \cdots \rangle_r$ denotes an average over each end state. $\mathcal{T} $ is then  calculated as the relative reduction in entropy between backwards and forwards path probability distributions, as given by Equation \ref{tree_eqn}.

\bibliographystyle{apsrev4-1}
%

\end{document}